\documentclass[showpacs,preprintnumbers,amsmath,amssymb]{elsarticle}

\usepackage{hyperref}
\usepackage{graphicx}
\usepackage{dcolumn}
\usepackage{bm}
\usepackage{amsmath}
\usepackage{slashed}
\usepackage{latexsym}
\usepackage{subfig}
\usepackage{multirow}

\newcommand{\bq}{\begin{equation}}
\newcommand{\eq}{\end{equation}}
\newcommand{\bqa}{\begin{eqnarray}}
\newcommand{\eqa}{\end{eqnarray}}
\newcommand{\nn}{\nonumber \\}

\def\be     {\begin{equation}}
\def\ee     {\end{equation}}
\def\bea        {\begin{eqnarray}}
\def\eea        {\end{eqnarray}}
\def\bnn    {\begin{eqnarray*}}
\def\enn    {\end{eqnarray*}}

\def\DD {{\mathcal{D}}}
\def\REE {{\mathrm{Re}}}
\def\IMM {{\mathrm{Im}}}

\journal{Nuclear Physics B}

\begin{document}

\begin{frontmatter}

\title{The Kondo effect revisited: RG-improved perturbation theory based on the Schwinger-boson representation}
\author{Jae-Ho Han$^{a}$, Minh-Tien Tran$^{b,c}$, and Ki-Seok Kim$^{a,d}$}
\address{
$^{a}$Department of Physics, POSTECH, Pohang, Gyeongbuk 790-784, Korea \\
$^{b}$Asia Pacific Center for Theoretical Physics (APCTP), POSTECH, Pohang, Gyeongbuk 790-784, Korea \\
$^{c}$Institute of Physics, Vietnam Academy of Science and Technology, 10 Dao Tan, Hanoi, Vietnam \\
$^{d}$Institute of Edge of Theoretical Science (IES), POSTECH, Pohang, Gyeongbuk 790-784, Korea
}

\begin{abstract}
Resorting to the Schwinger-boson representation for the description of a localized magnetic-impurity state, we develop an RG-improved (renormalization group) perturbation theory for the Kondo effect. This Schwinger-boson based RG-improved perturbation theory covers the whole temperature range from a decoupled local moment state to a local Fermi-liquid state through the crossover temperature regime, shown from the specific heat and spin susceptibility of the magnetic impurity. The Schwinger-boson based RG-improved perturbation theory makes the strong coupling fixed point at IR (infrared) accessible from the gaussian one at UV (ultraviolet) within the perturbation framework, regarded to be complementary to the Schwinger-boson based NCA (non-crossing approximation) self-consistent theory [Phys. Rev. Lett. {\bf 96}, 016601 (2006)]. The existence of the perturbatively accessible strong coupling fixed point implies the nature on the statistics of spinons, not determined by hands but chosen by the nature of strongly coupled systems.
\end{abstract}


\end{frontmatter}



\section{Introduction}

The Kondo effect is one of the most well-known strong coupling problems in condensed matter physics, ever since Kondo's explanation of conductance minimum in some metals with magnetic impurities \cite{Kondo original}. In a system showing the Kondo effect, a localized magnetic-impurity state is decoupled from dynamics of conduction electrons at high temperatures, while at low temperatures, they become strongly coupled and the local moment is screened completely by conduction electrons. In the low-temperature phase, the magnetic moment and conduction electrons form a many-body singlet state, referred to as a local Fermi-liquid phase \cite{Kondo_Textbook}. This strongly coupled dynamics has been solved by some exact methods such as numerical renormalization group \cite{NRG_Review} and Bethe ansatz \cite{Bethe_Ansatz}. They are non-perturbative in nature, which means that these theoretical frameworks need not assume any fixed points as a starting point. It would be valuable to develop a formalism which can describe the magnetic-impurity dynamics over the whole temperature range (from the decoupled local moment state as a gaussian UV (ultraviolet) fixed point to the local Fermi liquid state as an IR (infrared) strong coupling fixed point) and identify relevant quantum corrections responsible for the change from the UV to IR fixed points.
There is a sincere theory that describes the evolution of impurity dynamics from the local moment state to the local Fermi-liquid phase, referred to as self-consistent CTMA (conserving t-matrix approximation) theory, which is based on the slave-boson representation for the magnetic impurity \cite{CTMA_Review}. In this theory, renormalizations of interaction vertices are introduced into the slave-boson NCA (non-crossing approximation) theory. Such vertex corrections turn a weak coupling ``local" non-Fermi liquid fixed point of the slave-boson NCA theory into the strong coupling local Fermi-liquid state of the slave-boson CTMA theory. On the other hand, the Schwinger-boson based self-consistent theory has been demonstrated to make the local Fermi-liquid fixed point accessible from the local moment state within the NCA framework \cite{Schwinger_Boson_NCA}, regarded to be much simpler than the slave-boson CTMA framework.

In this paper we revisit the Kondo problem and confirm the existence of the perturbatively accessible strong coupling fixed point, starting from the decoupled local moment state. Resorting to the Schwinger-boson representation for the description of a localized magnetic-impurity state, we develop an RG-improved (renormalization group) perturbation theory for the Kondo effect. Evaluating the specific heat and spin susceptibility of the local magnetic moment, we show that the Schwinger-boson based RG-improved perturbation theory covers the whole temperature range from the decoupled local moment state to the local Fermi-liquid state through the crossover temperature regime. The Schwinger-boson based RG-improved perturbation theory makes the IR strong coupling fixed point accessible from the UV gaussian one within the perturbation framework, regarded to be complementary to the Schwinger-boson based NCA self-consistent theory \cite{Schwinger_Boson_NCA}. On the other hand, the slave-boson based RG-improved perturbation theory allows only the local non-Fermi liquid fixed point instead of the local Fermi-liquid state as the slave-boson NCA theory \cite{CTMA_Review}. This comparison leads us to speculate that the role of vertex corrections in the slave-boson CTMA theory is to change the statistics of spinons from fermions to bosons. We suggest that the statistics of spinons is not determined by hands but chosen by the nature of strongly coupled systems.

One may ask what we mean by ``perturbatively accessible strong coupling fixed point" \cite{Kim_Tanaka_Review}. A strong coupling fixed point means a fixed point defined at an infinite coupling constant. Here, the $\beta-$function for the Kondo coupling constant is negative, leading it to flow toward an infinite coupling fixed point at IR, identified with the local Fermi-liquid state. ``Perturbatively accessible" means that such a fixed point can be reached by incorporating relevant singular quantum corrections up to an infinite order, for example, in a way of the geometric sum. At first glance, the coexistence between ``perturbatively accessible" and ``strong coupling fixed point" looks incompatible. In order to resolve this inconsistency, suppose the BCS (Bardeen-Cooper-Schrieffer) problem on superconductivity \cite{BCS_Textbook}. The $\beta-$function for the effective pairing interaction constant is also negative, giving rise to a run-away flow at IR. However, we all know that this problem can be solved within the so called BCS mean-field theory, where the basis set is reconstructed, referred to as the Nambu spinor representation. On the other hand, one may introduce renormalizations for the effective pairing interaction coefficient into the BCS effective Hamiltonian via the Bethe-Salpeter equation and find the corresponding electron self-energy based on this singular interaction vertex, resulting in the same electron Green's function as the BCS one \cite{BCS_Textbook}. In this case the original normal electron basis has been utilized. In the single-impurity Kondo problem, we perform essentially the same work, where the Schwinger-boson representation is introduced to reconstruct the basis set for the magnetic impurity state and the renormalization group analysis is resorted to instead of the Bethe-Salpeter equation, both of which should be regarded to be identical at low energies.

The outline of this paper is as follows: In section \ref{RG eq} we perform the perturbative renormalization group analysis to derive the renormalized Kondo interaction vertex as a function of frequency. Then, using the frequency dependent coupling, we construct the RG-improved perturbation theory in Sec. \ref{RG-imp}. By evaluating the specific heat and spin susceptibility of the magnetic impurity, we show that the Schwinger-boson RG-improved perturbation theory can describe the whole temperature region from the decoupled local moment state to the local Fermi-liquid phase. In Section \ref{Discussion}, we discuss implications of our study, particularly, comparing our bosonic approach to the fermionic one such as the slave-boson based CTMA theory and give our speculations on strong coupling problems such as bad metal physics and heavy-fermion quantum criticality.

\section{Renormalization group of Kondo problem: Schwinger boson representation of impurity spin}
\label{RG eq}

\subsection{Kondo Hamiltonian with Schwinger boson representation of spin}

We start with the Kondo Hamiltonian given by
\begin{eqnarray}
H = \sum_{\vec k, \alpha} \epsilon_{\vec k} c_{\vec k \alpha}^\dagger c_{\vec k \alpha} + J_K \!\sum_{\vec k, \vec k',\alpha, \beta} \! \left(\frac{1}{2} c_{\vec k \alpha}^\dagger \vec \sigma_{\alpha \beta}  c_{\vec k' \beta} \right) \cdot \vec S,
\end{eqnarray}
where $\epsilon_{\vec k}$ is an energy of conduction electron with momentum $\vec k$, $c_{\vec k \alpha}^\dagger$ and $c_{\vec k, \alpha}$ are creation and annihilation operators of conduction electrons with momentum $\vec k$ and spin $\alpha=\uparrow, \downarrow$, respectively, $J_K$ is the Kondo coupling constant, $\vec \sigma = (\sigma^x, \sigma^y, \sigma^z)$ is a vector with Pauli matrices as components, and $\vec S$ is an impurity-spin operator. The first term is the Hamiltonian for single-band conduction electrons and the second term gives an interaction between the conduction electrons and impurity-spin at the origin. Here, we use the Schwinger-boson representation \cite{Spin_Textbook, SB_spin liquids} for an impurity-spin state $\vec S$
\begin{eqnarray}
S^+ &\equiv& S^x + iS^y = z_1^\dagger z_2, \ \ \
S^- \equiv S^x - iS^y = z_2^\dagger z_1, \ \ \ \nonumber \\
S^z &=& \frac{1}{2} \left( z_1^\dagger z_1 - z_2^\dagger z_2 \right)
\end{eqnarray}
with a constraint $z_1^\dagger z_1 + z_2^\dagger z_2 = 2S$, where $z_{1}$ and $z_{2}$ represent a doublet state of the impurity spin with an eigenvalue $S$ of $\vec S^2 = S(S+1)$. Then, the Kondo interaction term becomes
\begin{eqnarray}
H_K
&\equiv& J_K \!\sum_{\vec k, \vec k',\alpha, \beta} \! \left(\frac{1}{2} c_{\vec k \alpha}^\dagger \vec \sigma_{\alpha \beta}  c_{\vec k' \beta} \right) \cdot \vec S \nonumber \\
&=& \frac{J_K}{2} \!\sum_{\vec k, \vec k',\alpha, \beta} c_{\vec k \alpha}^\dagger c_{\vec k' \beta} z_{\beta}^\dagger z_{\alpha} - \frac{J_K}{4} \!\sum_{\vec k, \vec k',\alpha, \beta} c_{\vec k \alpha}^\dagger c_{\vec k' \alpha} z_{\beta}^\dagger z_{\beta}.
\label{eq:KondoInt}
\end{eqnarray}
where the identity  $\vec\sigma_{\alpha\beta} \cdot \vec\sigma_{\alpha' \beta'} = 2\delta_{\alpha\beta'}\delta_{\beta\alpha'} - \delta_{\alpha\beta}\delta_{\alpha'\beta'}$ has been used. In the second line of above equation, the second term induces a scattering without impurity-spin-flip which is the same as an elastic scattering with potential field. This is not important for the Kondo effect in single impurity problem, so we drop it and retain only the first term in this study.

In order to make our approximation controllable, it is natural to consider the $Sp(N)$ generalization for the Schwinger-boson representation \cite{SpN_Read_Sachdev}, given by
\begin{eqnarray}
H = \sum_{\vec k, \alpha} \epsilon_{\vec k} c_{\vec k \alpha}^\dagger c_{\vec k \alpha} + \frac{J_K}{N} \!\sum_{\vec k, \vec k',\alpha, \beta} c_{\vec k \alpha}^\dagger c_{\vec k' \beta} z_{\beta}^\dagger z_{\alpha},
\label{eq:Hamiltonian}
\end{eqnarray}
where the indices $\alpha, \beta = 1, 2$ are extended to have values $\alpha, \beta = 1, 2, ..., N$ and the Kondo coupling constant redefined as $J_K/2 \rightarrow J_K/N$ so that the coupling term has an order of $N$ rather than $N^2$. The constraint of $z$-fields is generalized to $\sum_\alpha z_\alpha^\dagger z_\alpha = NS$. Then, quantum corrections can be suppressed by the control parameter $1/N$ in the $N \rightarrow \infty$ limit.

We would like to point out that it is essential to introduce the channel number $K$ of conduction electrons in the Schwinger-boson NCA framework, classifying the Kondo effect into the exactly screened case $\gamma \equiv K / N = S$, the over-screened case $\gamma > S$, and the under-screened case $\gamma < S$ within the self-consistent analysis \cite{Schwinger_Boson_NCA}. On the other hand, our RG-improved Schwinger-boson perturbation theory does not care about the channel number of conduction electrons since it is not self-consistent but RG-improved perturbative, where the information of the RG analysis determines the nature of the Kondo effect. We emphasize that this does not mean inconsistency between the self-consistent Schwinger-boson NCA theory and the RG-improved Schwinger-boson perturbation framework. Instead, it implies that it is not clear how the channel number of conduction electrons can be encoded into the RG analysis consistently. In Sec. $\ref{Discussion}$ we discuss how the over-screened Kondo effect can be achieved within the RG-improved Schwinger-boson perturbation theory.

%
%

\subsection{Setup for the renormalization group, scaling analysis}

We write down the partition function for the Kondo problem of Eq. (\ref{eq:Hamiltonian}) (in the imaginary time formulation) as follows
\begin{eqnarray}
Z = \int\!\! \DD c \DD z \ e^{-S[c, z; \lambda]},
\end{eqnarray}
where
\begin{eqnarray}
S[c, z; \lambda]
&=& \int_0^\beta\!\! d\tau \ \Bigg[ \sum_{\vec k, \alpha} c_{\vec k \alpha}^\dagger (\tau) \partial_\tau c_{\vec k \alpha}(\tau) + \sum_\alpha z_{\alpha}^\dagger(\tau) \partial_\tau z_{\alpha}(\tau)  \nn
&& + \sum_{\vec k, \alpha} c_{\vec k \alpha}^\dagger (\tau) \big( \epsilon_{\vec k} -\mu \big) c_{\vec k \alpha}(\tau) + \frac{J_K}{N} \!\! \sum_{\vec k, \vec k', \alpha, \beta} \!\! c_{\vec k \alpha}^\dagger(\tau) c_{\vec k' \beta}(\tau) z_{\beta}^\dagger(\tau) z_{\alpha}(\tau) \nn
&&\hspace{120pt} + \lambda \Big( \sum_{\alpha} z_{\alpha}^\dagger(\tau) z_{\alpha}(\tau) - N S \Big) \Bigg], \label{eq:Starting_Action}
\end{eqnarray}
where $\mu$ is chemical potential for conduction electrons.
The last term comes from the constraint in the $Sp(N)$ generalization, and $\lambda$ is a Lagrange multiplier. We will use saddle-point approximation for $\lambda$ : $\partial_\lambda F[\lambda]=0$ and $F[\lambda] = -\frac{1}{\beta} \log Z$. It plays the role of a chemical potential in dynamics of Schwinger bosons, where its finite value at zero temperature turns out to be essential for the emergence of the local Fermi-liquid state from the decoupled local moment state.

For a scaling analysis, it is convenient to express the action in the frequency and momentum space:
\begin{eqnarray}
S[c, z; \lambda]
&=& \sum_{\alpha} \int\!\! \frac{d\omega}{2\pi} \frac{d^3k}{(2\pi)^3} \ c_{\alpha}^\dagger (i\omega,\vec k) \big( -i\omega + \xi_{\vec k} \big) c_{\alpha}(i\omega, \vec k) \nn
&+& \sum_\alpha \int\!\! \frac{d\Omega}{2\pi} \ z_{\alpha}^\dagger(i\Omega) \big( -i\Omega + \lambda \big) z_{\alpha}(i\Omega) \nn
&+& \frac{J_K}{N} \sum_{\alpha, \beta} \int\!\! \frac{d\omega}{2\pi} \frac{d\omega'}{2\pi} \frac{d\nu}{2\pi} \frac{d^3k}{(2\pi)^3} \frac{d^3k'}{(2\pi)^3} \nn
&&\hspace{20pt}\times c_{\alpha}^\dagger(i\omega, \vec k) c_{\beta}(i\omega', \vec k') z_{\beta}^\dagger(i\omega'+i\nu) z_{\alpha}(i\omega + i\nu)  - \beta\lambda NS .
\label{eq:Action_Engy}
\end{eqnarray}
Here, we considered the Fourier transformation of $c_{\vec k \alpha}(\tau) = \frac{1}{\sqrt{V}} \int\!\! \frac{d\omega}{2\pi} \ e^{-i\omega \tau} c_{\alpha}(i\omega, \vec k)$ and $z_\alpha(\tau) = \int\!\! \frac{d\Omega}{2\pi} \ e^{-i\Omega \tau} z_{\alpha}(i\Omega)$. The existence of a Fermi surface gives rise to anisotropic scaling of the momentum: only the momentum component perpendicular to the Fermi surface scales, not the parallel components \cite{RG_Shankar_RMP}. Then, we consider the following scale transformation,
\begin{eqnarray}
k_\parallel &=& \frac{k_\parallel'}{b}, \ \ \ \vec k_\perp = \vec k_\perp', \ \ \ i\omega = \frac{i\omega'}{b},
\end{eqnarray}
where $k_\parallel$ ($\vec k_\perp$) is the momentum component parallel (perpendicular) to normal vector of the Fermi surface and $b>1$ is a scaling factor. As our reference, the scale transformation
\begin{eqnarray}
c_\alpha(i\omega, \vec k) = b^{\frac{3}{2}} c_\alpha' (i\omega', \vec k'), \ \ \ z_\alpha (i\omega) = b z_\alpha' (i\omega').
\end{eqnarray}
lead each kinetic-energy term invariant. Based on this scaling transformation, we find that the Kondo coupling is marginal:
\begin{eqnarray}
J_K' = J_K.
\end{eqnarray}

The above discussion allows us to introduce the ``dimensional" regularization scheme, modifying the density-of-states (DOS) of conduction electrons as
\begin{eqnarray}
\rho_c(E)
&=& \rho_0, \nonumber \\
&\rightarrow&
\rho_0 |E|^{-\epsilon}
\label{eq:Reg}
\end{eqnarray}
where $E$ is the energy of conduction electrons from the chemical potential $\mu$ and $\epsilon$ is a positive infinitesimal number as a regulator. The modification of DOS is equivalent to extending the dimension of perpendicular to the Fermi surface: $\int dk_\parallel \rightarrow \int d^{1+\epsilon} k_\parallel$ and $E(k) = v_F k_\parallel \rightarrow v_F k^{1+\epsilon}_\parallel$. Accordingly, the scaling factors of the fields change as
\begin{eqnarray}
c_\alpha(i\omega, \vec k) = b^{\frac{3}{2} + \frac{\epsilon}{2}} c_\alpha' (i\omega', \vec k'), \ \ \ z_\alpha (i\omega) = b z_\alpha' (i\omega'),
\end{eqnarray}
and the Kondo coupling scales as
\begin{eqnarray}
J_K' = b^{-\epsilon} J_K .
\end{eqnarray}
%
%

Separating the action into the renormalized and counter parts,
\begin{eqnarray}
S = S_r + S_c - \beta\lambda_rNS ,
\end{eqnarray}
we obtain
\begin{eqnarray}
S_r
&=& \sum_{\alpha} \int_k \ c_{\alpha}^\dagger \big( -i\omega + \xi_{\vec k} \big) c_{\alpha} + \sum_\alpha \int_\Omega \ z_{r,\alpha}^\dagger \big( -i\Omega + \lambda_r \big) z_{r, \alpha} \nn
&+& \frac{J_{K,r}}{N}\mu^\epsilon \sum_{\alpha, \beta} \int_{k, k', \nu} c_{\alpha}^\dagger c_{\beta} z_{r,\beta}^\dagger z_{r,\alpha}, \\
S_c
&=& \sum_\alpha \int_\Omega \ (Z_z-1) z_{r,\alpha}^\dagger \big( -i\Omega \big) z_{r, \alpha} + (Z_\lambda-1) \lambda_r z_{r,\alpha}^\dagger z_{r, \alpha}\nn
&+& (Z_J - 1)\frac{J_{K,r}}{N} \mu^\epsilon \sum_{\alpha, \beta} \int_{k, k', \nu} c_{\alpha}^\dagger c_{\beta} z_{r,\beta}^\dagger z_{r,\alpha}.
\end{eqnarray}
Here the subscript $r$ indicates renormalized quantities, $Z$'s are renormalization factors, and the integrals are abbreviated and the arguments of the fields are suppressed for simplicity. A `mass' scale $\mu$ is introduced to make the coupling constant $J_{K,r}$ be dimensionless. The relation between the renormalized and bare quantities are \cite{RG_Textbook}
\begin{eqnarray}
z_{\sigma} = Z_{z}^{\frac{1}{2}} z_{r\sigma} , \ \ \
\lambda = Z_{z}^{-1} Z_{\lambda} \lambda_{r} , \ \ \
J_K = Z_{z}^{-1} Z_{J} \mu^{\epsilon} J_{K,r}.
\end{eqnarray}
Note that the conduction electrons are assumed not to be renormalized in the single-impurity Kondo problem. The Green's functions or propagators of electrons and $z$-field (spinons) are
\begin{eqnarray}
\mathcal{G}_c^{(0)}(i\omega, \vec k)
&=& \frac{1}{i\omega + \mu - \epsilon_{\vec k}}, \\
\mathcal{G}_z^{(0)}(i\Omega)
&=& \frac{1}{i\Omega - \lambda_r},
\end{eqnarray}
respectively. Based on these building blocks, we evaluate self-energy and vertex corrections and find their singular parts.

\subsection{Vertex corrections up to one-loop order}

Here after, we use summation of frequency and momentum instead of integral, $\int\!\!\frac{d\omega}{2\pi} \rightarrow \frac{1}{\beta}\sum_{i\omega}$ and $\int\!\! \frac{d^dk}{(2\pi)^3} \rightarrow \sum_{\vec k}$, for computational convenience.

In the zeroth order, the four-point vertex function is just the coefficient of the interaction term:
\begin{eqnarray}
\Gamma^{(0)} (i\nu)
&=& \frac{J_{K,r}}{N\beta} \mu^{\epsilon}.
\end{eqnarray}
%
%
%
%
\begin{figure*}[t!]
\includegraphics[width=12cm]{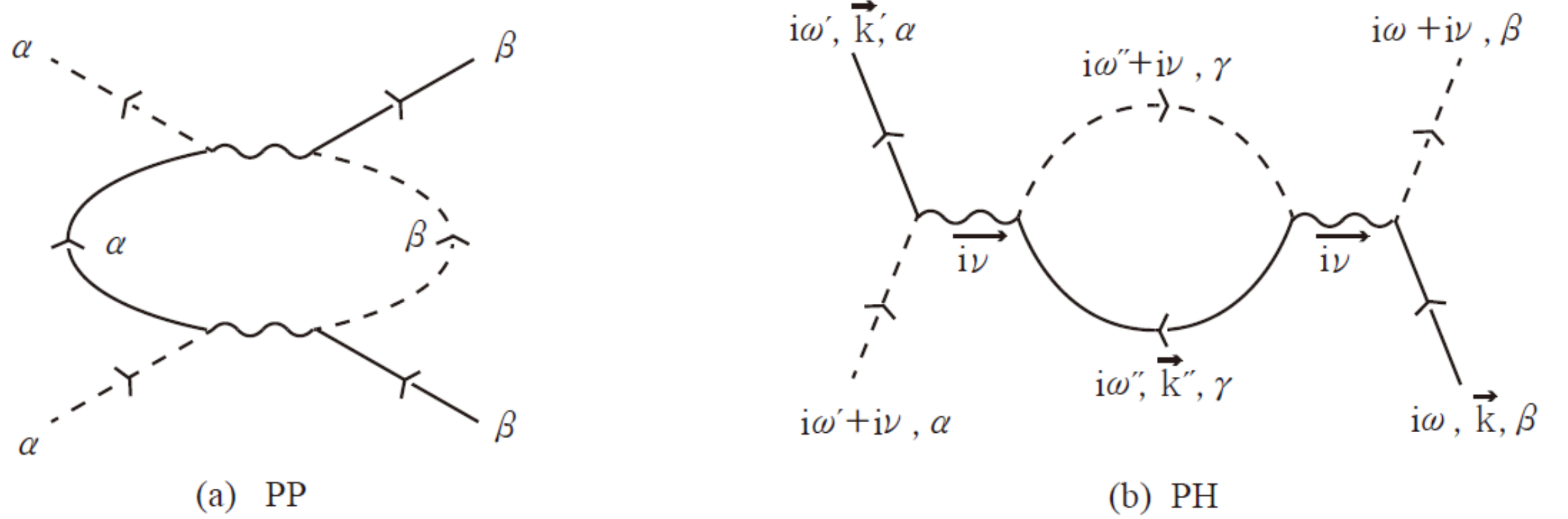}
\caption{
(a) Particle-Particle (PP) and (b) Particle-Hole (PH) diagrams for the four-point vertex function in the one-loop level.
}
\label{fig_1loop_vertex}
\end{figure*}
%
%
In the one-loop order, there are two kinds of diagrams, which correspond to particle-particle (PP) and particle-hole (PH) channels (Fig. \ref{fig_1loop_vertex}). Considering the order in $1/N$ of each graph, the PP term is in the order of $1/N^2$ while the PH term is in the order of $1/N$. As a result, we retain only the PH term, given by
\begin{eqnarray}
\Gamma_{PH}^{(1)}(i\nu)
= -\frac{J_{K,r}^2}{N\beta^2}\mu^{2\epsilon} \sum_{i\omega'', \vec k''} \ \mathcal{G}_c^{(0)}(i\omega'',\vec k'') \mathcal{G}_z^{(0)}(i\omega''+i\nu).
\end{eqnarray}
The frequency summation can be easily performed. Converting the $\vec k''$-summation into an energy integral, $\sum_{\vec k''} \rightarrow \int\!\! dE \ \rho_c(E)$, and considering the zero-temperature limit ($f(E) \rightarrow \Theta(-E)$, where $f(E)$ is the Fermi-Dirac distribution function), the above expression is
\begin{eqnarray}
\Gamma_{PH}^{(1)}(i\nu)
&=& -\frac{J_{K,r}^2}{N\beta} \mu^{2\epsilon} \rho_0 \int_{-\infty}^{0}\!\! dE \ \frac{|E|^{-\epsilon}}{i\nu - \lambda_r + E} \nn
&=& \frac{J_{K,r}^2}{N\beta} \mu^{2\epsilon} \rho_0 \pi \left( -i\nu + \lambda_r \right)^{-\epsilon} \csc (\pi\epsilon) \nn
&=& \frac{J_{K,r}}{N\beta}\mu^{\epsilon} \left\{ \frac{\rho_0 J_{K,r} }{\epsilon} - \rho_0 J_{K,r} \log\left( \frac{-i\nu + \lambda_r}{\mu} \right) \right\} + \mathcal{O}(\epsilon).
\end{eqnarray}
The divergent term proportional to $1/\epsilon$ is canceled, adding the counter term of $\Gamma_C = \frac{J_{K,r}}{N}\mu^{\epsilon} (Z_J - 1)$ if
\begin{eqnarray}
Z_J = 1 - \rho_0 J_{K,r} \frac{1}{\epsilon}
\end{eqnarray}
is satisfied. This relation is one of the equations needed to derive RG equations in later subsection.

\subsection{Self-energy corrections up to two-loop order}

\begin{figure*}[t!]
\includegraphics[width=13cm]{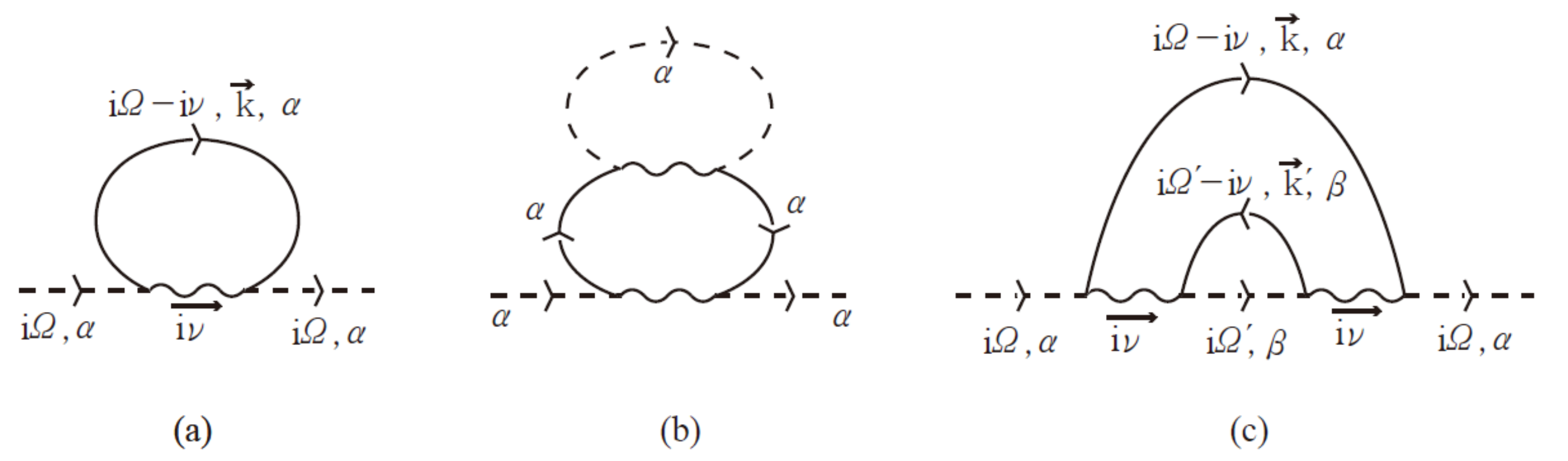}
\caption{
(a) One-loop self-energy correction and two-loop self-energy corrections proportional to (b) $1/N^2$ and (c) $1/N$.
}
\label{fig_loop_self_energy}
\end{figure*}

The one-loop spinon self-energy is (Fig. \ref{fig_loop_self_energy} (a))
\begin{eqnarray}
\Sigma_z^{(1)}(i\Omega)
&=& -\frac{J_{K,r}}{N\beta}\mu^{\epsilon} \sum_{\vec k'', i\nu} \mathcal{G}_c^{(0)}(i\Omega-i\nu, \vec k'').
\end{eqnarray}
Repeating the same procedure as the case of the vertex function, we obtain
\begin{eqnarray}
\Sigma_z^{(1)}(i\Omega)
&=& \frac{J_{K,r}}{N}\mu^{\epsilon} \rho_0 \int_{0}^{D} \!\! dE \ |E|^{-\epsilon} \nn
&=& \frac{J_{K,r}}{N} \rho_0 D + \mathcal{O}(\epsilon).
\end{eqnarray}
where $D$ is half of the band width. Here, the divergence (in $D\rightarrow \infty$ limit) is independent of $1/\epsilon$ and can be absorb into the chemical potential $\lambda$. Therefore it does not influence the RG equations. To find a singular part ($1/\epsilon$-divergence), consider two-loop corrections.
%
%
%
%

There are two kinds of two-loop self-energy diagrams, but again considering the order in $1/N$ for the diagrams, the only surviving contribution is given by Fig. \ref{fig_loop_self_energy}c. The expression for this diagram is
\begin{eqnarray}
\Sigma_z^{(2)}(i\Omega)
= -\frac{J_{K,r}^2}{N\beta^2} \mu^{2\epsilon} \!\!\! \sum_{i\nu, i\Omega', \vec k, \vec k'} \!\!\! \mathcal{G}_c^{(0)}(i\Omega-i\nu, \vec k) \mathcal{G}_c^{(0)}(i\Omega'-i\nu, \vec k') \mathcal{G}_z^{(0)}(i\Omega').
\end{eqnarray}
As before, performing frequency summations first and converting momentum summations into energy integrals, we obtain
\begin{eqnarray}
\Sigma_z^{(2)}(i\Omega)
&=& -\frac{J_{K,r}^2}{N} \mu^{2\epsilon} \rho_0^2 \int_0^\infty \!\!dE dE' \ \frac{E^{-\epsilon} E'^{-\epsilon}}{E + E' + \lambda_r - i\Omega} \nn
&=& \frac{J_{K,r}^2}{N} \rho_0^2 (-i\Omega + \lambda_r) \Bigg[ \frac{1}{2\epsilon} + 1 + \log\left( \frac{-i\Omega + \lambda_r}{\mu} \right) \Bigg] + \mathcal{O}(\epsilon).
\end{eqnarray}
in the zero-temperature limit. Here, the $1/\epsilon$ divergence can be canceled, adding the counter term of $\Sigma_{z,C} = (Z_z-1)(-i\Omega) + (Z_\lambda-1)\lambda_r$, if
\begin{eqnarray}
Z_z = Z_\lambda = 1 - \frac{\rho_0^2 J_{K,r}^2}{2N} \frac{1}{\epsilon}
\end{eqnarray}
are satisfied.

\subsection{Renormalization group equations and effective Kondo coupling}

Recalling the relation between bare and renormalized parameters of $\lambda = Z_z^{-1} Z_\lambda \lambda_r$ and $J_K = Z_z^{-1} Z_J \mu^{\epsilon} J_{K,r}$
with $Z_J = 1 - \rho_c J_{K,r} \frac{1}{\epsilon}$ and $Z_z = Z_\lambda = 1 - \frac{\rho_0^2 J_{K,r}^2}{2N} \frac{1}{\epsilon}$, and considering the independence of bare parameters in
the energy scale of $\mu$, i.e., ${d\lambda}/{d\mu} = 0$ and ${dJ_K}/{d\mu}=0$, we obtain the following RG equations (with $\epsilon \rightarrow 0$)
\begin{eqnarray}
\frac{d\lambda_r}{d\log\mu} = 0, \ \ \
\frac{dJ_{K,r}}{d\log\mu} = - \rho_0 J_{K,r}^2 + \frac{\rho_0^2}{N} J_{K,r}^3.
\label{eq:RGE}
\end{eqnarray}
When neglecting the second (cubic) term in the $J_{K,r}$-equation, we have exactly screened Kondo effect, while retaining it we have the over-screened Kondo effect. Here, we will consider the exactly screened Kondo effect and neglect the second term. Then we find an effective chemical potential and a renormalized Kondo coupling constant as a function of $\mu$
\begin{eqnarray}
\lambda_r(\mu) = \mathrm{const.} \equiv \lambda_D, \ \ \ J_{K,r}(\mu) = \frac{1}{\rho_0 \log \frac{\mu}{T_K}},
\end{eqnarray}
respectively, where the Kondo temperature $T_K$ is defined by $T_K = D \exp \left[ -\frac{1}{\rho_0 J_{K,r}(D)} \right]$.

The $\beta-$function for the Kondo coupling constant is the same as that of the poor man's or woman's scaling analysis, where renormalizations for wave functions are not taken into account.
In addition, this $1/N$ result is consistent with the $t-$matrix approximation based on the Bethe-Salpeter equation for the Kondo coupling constant as the case of the BCS theory. This RG result
is not modified even if we resort to the slave-boson representation instead of the Schwinger-boson basis, where spinons are fermions.

\section{Renormalization-improved perturbation theory}
\label{RG-imp}

\subsection{An effective action with a scale-dependent renormalized coupling constant}

Incorporating the energy dependence of the Kondo coupling into Eq. (\ref{eq:Starting_Action}), we write down an effective low-energy action
\begin{eqnarray}
S_r[c, z;\lambda]
&=& \sum_{i\omega, \vec k, \alpha} c_{\vec k \alpha}^\dagger (i\omega) \big(-i\omega - \mu + \epsilon_{\vec k} \big) c_{\vec k \alpha}(i\omega) \nn
&+& \sum_{i\Omega, \alpha} z_{r\alpha}^\dagger (i\Omega) \big( -i\Omega + \lambda_D \big) z_{r\alpha}(i\Omega) \nn
&+& \frac{1}{N \beta V} \sum_{\substack{i\nu, i\omega, i\omega', \\ \vec k, \vec k', \alpha, \beta }} J(i\nu) c_{\vec k' \alpha}^\dagger (i\omega') c_{\vec k \beta}(i\omega) z_{r\beta}^\dagger (i\omega + i\nu) z_{r\alpha} (i\omega'+ i\nu) \nn
&-& \beta \lambda_D NS, \label{RG_Effective_Action}
\end{eqnarray}
where
\begin{eqnarray}
J(i\nu) = \frac{1}{\rho_c \log\frac{i\nu}{T_K}}.
\end{eqnarray}

We would like to emphasize that Eq. (\ref{RG_Effective_Action}) differs from Eq. (\ref{eq:Starting_Action}) in the respect of field contents. Eq. (\ref{RG_Effective_Action}) is defined in the energy scale of $\mu$ much lower than the original cutoff scale. Thus, both conduction electrons and bosonic spinons should be regarded as emergent renormalized particles at low energies, interacting with each other though renormalized (nonlocal in time) effective Kondo interactions. In this respect quantum corrections with effective Kondo interactions for the dynamics of these low-energy excitations should be distinguished from those of the RG analysis.

\subsection{Low energy dynamics of spinons}

The one-loop spinon self-energy is the same as before formally but with $J \rightarrow J(i\nu)$:
\begin{eqnarray}
\Sigma_{z} (i\Omega)
&=& -\frac{1}{N\beta} \sum_{i\nu, \vec k} J(i\nu) \mathcal{G}_{c}^{(0)} (i\Omega - i\nu) \nn
&=& -\frac{1}{N\beta} \sum_{i\nu, \vec k} \frac{1}{\rho_c \log\frac{i\nu}{T_K}} \frac{1}{i\Omega - i\nu - \xi_{\vec k}}.
\label{eq:self_energy}
\end{eqnarray}
Observing that all singularities of the interaction vertex $J(z)=\frac{1}{\rho_c \log\frac{z}{T_K}}$ lie along the real axis (a pole at $z=T_K$ and a branch cut along the negative real axis), we can deform a contour to allow the following expression,
\begin{eqnarray}
J(i\nu)
= -\frac{1}{\pi} \int_{-\infty}^{\infty}\!\!dE \ \frac{1}{i\nu - E} \IMM \left\{ \frac{1}{\rho_c\log\frac{E + i\delta}{T_K}} \right\}.
\end{eqnarray}
Inserting this expression into Eq. (\ref{eq:self_energy}) and performing the frequency summation, we obtain
\begin{eqnarray}
\Sigma_{z} (i\Omega)
= -\frac{1}{\pi \rho_c N} \sum_{\vec k} \int_{-\infty}^{\infty}\!\!dE \ \frac{f(E) - f(\xi_{\vec k})}{i\Omega + E - \xi_{\vec k}} s \IMM \left\{ \frac{1}{\log\frac{E+i\delta}{T_K}} \right\}.
\end{eqnarray}
Using $\IMM \left\{ 1/\log\frac{E+i\delta}{T_K} \right\} = -\pi / \left\{ \left( \log\frac{|E|}{T_K} \right)^2 + \pi^2 \right\}$ when $E<0$ and taking the $T\rightarrow 0$ limit (so that $f(E) \rightarrow \Theta (-E)$), we obtain
\begin{eqnarray}
\Sigma_{z} (i\Omega)
= \frac{1}{N} \int_{0}^{\infty}\!\!dE dE' \frac{1}{i\Omega - E - E'} \frac{1}{\left( \log\frac{E}{T_K} \right)^2 + \pi^2}.
\end{eqnarray}

It is convenient to consider the retarded self-energy which can be obtained by the analytic continuation:
\begin{eqnarray}
\Sigma_{z}^R (\Omega)
&=& \Sigma_{z}(i\Omega \rightarrow \Omega + i\delta) \nn
&=& \frac{1}{N} \int_{0}^{\infty}\!\!dE dE' \frac{1}{\Omega - E - E' + i\delta} \frac{1}{\left( \log\frac{E}{T_K} \right)^2 + \pi^2}.
\end{eqnarray}
Recalling $\frac{1}{x+i\delta} = \mathcal{P}\frac{1}{x} - i\pi\delta(x)$, where $\mathcal P$ is Cauchy's principal value, the imaginary part of the retarded self-energy is
\begin{eqnarray}
\IMM\Sigma_{z}^R (\Omega)
&=& -\frac{\pi T_K}{N} \int_{0}^{-\frac{\Omega}{T_K}}\!\! dx \ \frac{\theta(-\Omega)}{\left( \log x \right)^2 + \pi^2}. \label{Imaginary_Self_Energy}
\end{eqnarray}
The real part of the retarded self-energy can be obtained, resorting to the Kramers-Kronig relation,
\begin{eqnarray}
\REE\Sigma_z^R (\Omega)
= \frac{1}{\pi} {\mathcal P} \int_{-\infty}^{\infty}\!\! d\Omega' \ \frac{\IMM\Sigma_z^R(\Omega')}{\Omega' - \Omega}.
\end{eqnarray}

\begin{figure}[t!]
\includegraphics[width=8cm]{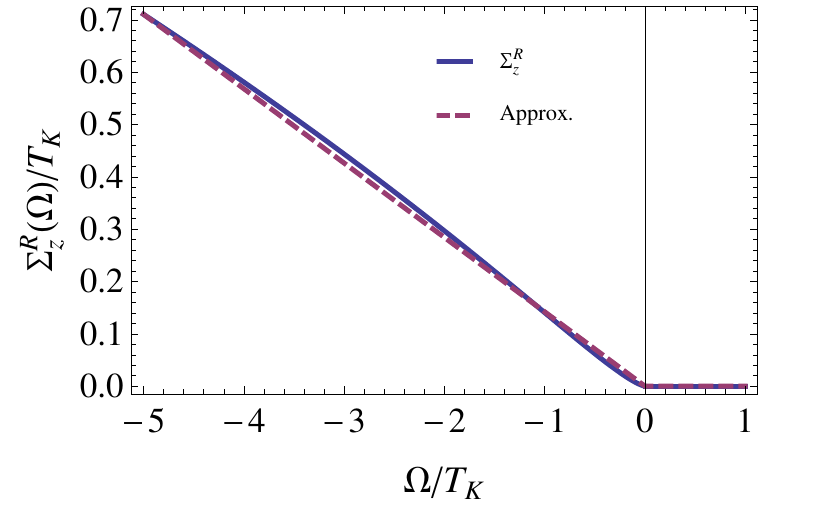}
\caption{
Imaginary part of the retarded self-energy (Eq. (\ref{Imaginary_Self_Energy})). The frequency-linear dependence shows marginal Fermi-liquid physics for the spinon dynamics,
originating from the branch-cut singular nature of the renormalized Kondo interaction vertex in the RG-improved perturbation theory.
}
\label{fig_Im_Sigma}
\end{figure}

The imaginary part of the retarded self-energy $\Sigma_z^R(\Omega)$ is shown in Fig. \ref{fig_Im_Sigma}.
%
%
%
%
The frequency dependence of $\IMM\Sigma_z^R(\Omega)$ can be fitted as a linear function, given by $\Sigma_z^R(\Omega) \approx - A \Omega \ \Theta(-\Omega)$ with $A=0.142$.
See the comparison between Eq. (\ref{Imaginary_Self_Energy}) and $\Sigma_z^R(\Omega) \approx - A \Omega \ \Theta(-\Omega)$ in Fig. \ref{fig_Im_Sigma}. Then, the real part is
\begin{eqnarray}
\REE\Sigma_z^R (\Omega)
&=& \frac{1}{\pi} {\mathcal P} \int_{-\infty}^{\infty}\!\! d\Omega' \ \frac{\IMM\Sigma_z^R(\Omega')}{\Omega' - \Omega} \nn
&=& \frac{1}{\pi} {\mathcal P} \int_{-\infty}^{0}\!\! d\Omega' \ \frac{-A \Omega'}{\Omega' - \Omega} \nn
&=& -\frac{A}{\pi} \Omega \log |\Omega| + C,
\label{eq_ReSigma}
\end{eqnarray}
where $C$ is an infinite constant. Such an infinite constant can be absorbed into the effective spinon chemical potential $\lambda_{D}$, which does not play any roles in the spinon dynamics.

Introducing this spinon self-energy into the impurity free energy of $F_{imp} = \frac{N}{\beta} \sum_{i\Omega} \log \big( -i\Omega + \lambda_D + \Sigma_z (i\Omega) \big) - \lambda_D NS$ and minimizing it with respect of $\lambda_{D}$, i.e., $\frac{\partial}{\partial \lambda_D} F_{imp} = 0$, the effective spinon chemical potential is determined by
\begin{eqnarray}
\frac{1}{\beta} \sum_{i\Omega} \frac{1}{-i\Omega + \lambda_D + \Sigma_z (i\Omega)} = S,
\end{eqnarray}
or
\begin{eqnarray}
&&\frac{1}{\pi} \int_{-\infty}^{\infty}\!\! dE \ n_B(E)  \IMM\left\{ \frac{1}{-E + \lambda_D + \Sigma_z^R (E) - i\delta} \right\}
= S.
\label{eq_lambda}
\end{eqnarray}
An essential point is that the spinon chemical potential remains finite even at zero temperature, which should be regarded to be a distinguished feature, compared with the fermion representation for spinons. In the fermion representation, the chemical potential of spinons should be zero to satisfy the half-filled condition of the Kondo regime  \cite{Kondo_Textbook}. In our case, even if the spinon chemical potential is finite at zero temperature, the emergence of huge damping
linearly proportional to frequency allows us to make the constraint of the Kondo regime satisfied.

%
%
\begin{figure}[t!]
\includegraphics[width=8cm]{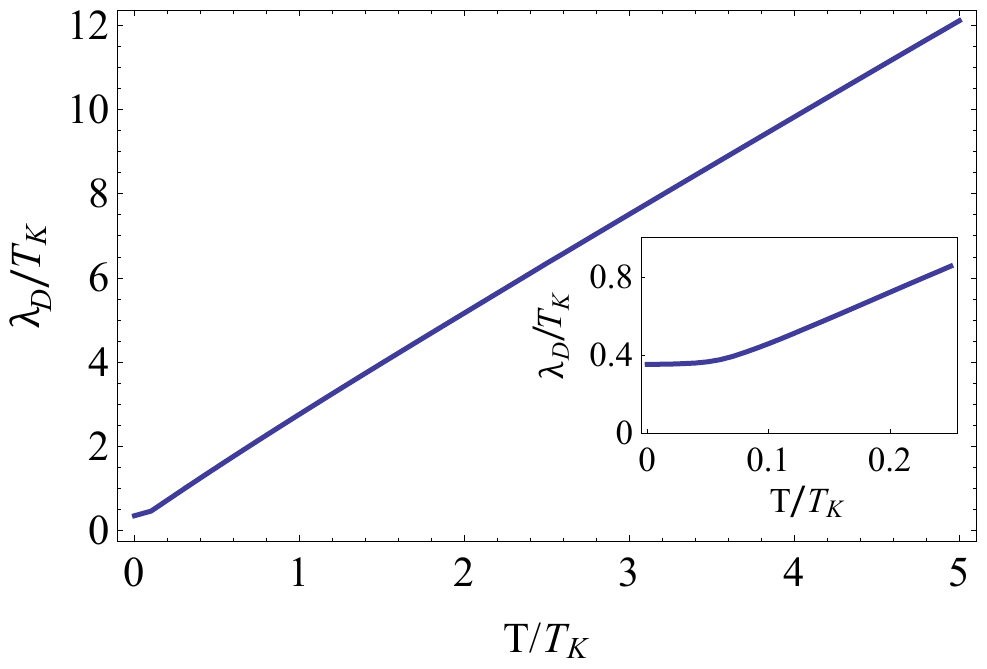}
\caption{Effective spinon chemical potential $\lambda_D (T)$ as a function of temperature. It is finite at zero temperature,
reflecting the aspect of the overdamped spinon dynamics (marginal Fermi-liquid phenomenology) and the bosonic nature of spinons.}
\label{fig_lambda}
\end{figure}
%
%

\begin{figure}[t!]
\includegraphics[width=8cm]{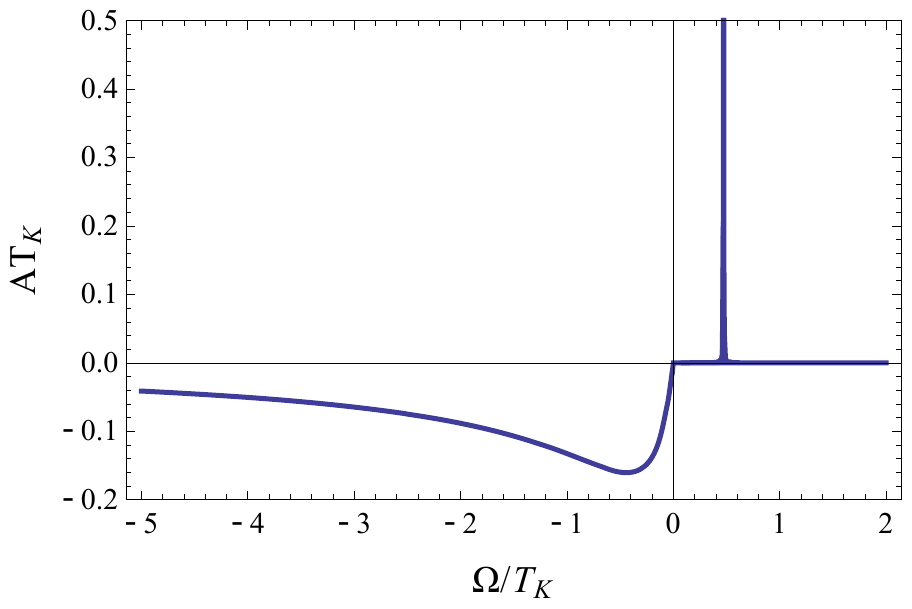}
\caption{
Typical form of the spectral function of spinons, plotted at $T/T_K = 0.1$. The positive-frequency part describes the decoupled local moment state while the negative-energy sector explains the strong-coupling local Fermi-liquid state.
}
\label{fig_spectlF}
\end{figure}

Introducing the spinon self-energy and effective spinon chemical potential into the spinon Green's function, we obtain the spectral function of spinons, given by
\begin{eqnarray}
\mathcal{A}(\Omega)
&=& -2 \IMM \ \mathcal{G}_z (i\Omega \rightarrow \Omega + i\delta) \nn
&=& -2 \IMM \ \frac{1}{\Omega - \Sigma^R_z (\Omega) + i\delta} \nn
&=& \left\{ \begin{array}{ll}
2\pi \delta (\Omega - \REE \Sigma^R_z(\Omega)), & \Omega >0, \\
\frac{-2 \IMM\Sigma^R_z(\Omega)}{ \left\{ \Omega - \REE\Sigma^R_z(\Omega) \right\}^2 + \left\{ \IMM\Sigma^R_z(\Omega) \right\}^2 }, & \Omega <0.
\end{array}\right.
\end{eqnarray}
See Fig. \ref{fig_spectlF}. The positive-frequency part describes the spectral intensity of the decoupled local moment state while the negative-frequency sector explains that of the strong coupling Kondo regime of the local Fermi-liquid state. It is essential to notice that the imaginary part of the spinon self-energy is linearly proportional to frequency, regarded to be immensely overdamped. Of course, this huge damping originates from strong Kondo interactions between localized spinons and conduction electrons, described by the renormalized effective Kondo vertex. Sometimes, the frequency-linear behavior of the self-energy is regarded to be a signature of a ``marginal" Fermi-liquid state, proposed to be a phenomenological theory for a non-Fermi liquid state near quantum criticality \cite{Varma_Review}. The emergence of the overdamped dynamics of spinons allows the excitation gap for the spinon dynamics to satisfy the number-constraint equation, where the excitation gap originates from the finite chemical potential at zero temperature. However, it generates an energy scale, below which the overdamped spinon dynamics gives rise to Fermi-liquid physics, identified with the local Fermi-liquid state. This aspect will be discussed later, based on the specific heat and spin susceptibility of the magnetic impurity, where the saturation behavior involved with the Kondo temperature starts to appear around this energy scale. Of course, this energy scale seems to coincide with that of the Schwinger-boson based NCA self-consistent theory \cite{Schwinger_Boson_NCA}.

%
%
%
%

Unfortunately, there exists one difficulty in this Schwinger-boson based RG-improved perturbation theory. When we find the effective spinon chemical potential, we set $N=2$ and $S=1$ instead of $S=1/2$.
We emphasize that this does not originate from our calculational mistakes. There is a solution for $\lambda_{D}$ even if $S = 1/2$ is considered. But, it turns out to be too large to describe the Kondo effect.
Comparing the Schwinger-boson based RG-improved perturbation theory with the Schwinger-boson based NCA self-consistent theory, where the latter seems to produce a reasonable energy scale for the Kondo effect with $S = 1/2$,
we could find that the overdamped spinon dynamics due to the renormalized singular Kondo vertex is overestimated in the RG-improved perturbation theory. As a result, the spinon spectral weight of the negative-frequency side in the Schwinger-boson based RG-improved perturbation theory is taken into account to be more than that in the Schwinger-boson based NCA self-consistent theory. In spite of this difficulty, we would like to emphasize
that the energy scale for the local Fermi-liquid state $T_{LFL}$, determined by the saturation of the spinon chemical potential, is consistent with that of the NCA result, given by $T_{LFL} / T_{K} \sim 0.1$.

\subsection{Thermodynamic quantities: specific heat ant spin susceptibility}

\begin{figure}[t!]
\includegraphics[width=8cm]{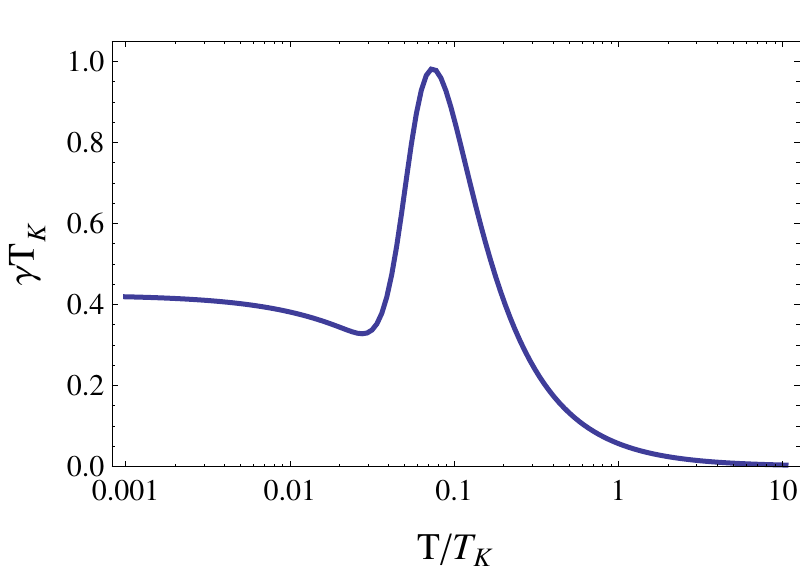}
\caption{ Specific heat coefficient $\gamma$ of a magnetic impurity in a log-normal scale. It shows three temperature regions, identified with the decoupled local moment state at high temperatures,
a Kondo fluctuating crossover regime at intermediate temperatures, and the local Fermi-liquid state at low temperatures. A dip should be regarded as an artifact
of the Schwinger-boson based RG-improved perturbation theory. See the text for details.
}
\label{fig_SpHeatCoeff}
\end{figure}

The specific heat coefficient of the magnetic impurity is given by
\begin{eqnarray}
\gamma = \frac{dS_{imp}}{dT} ,
\end{eqnarray}
where the impurity-spin entropy is
\begin{eqnarray}
S_{imp} = -\frac{\partial F_{imp}}{\partial T} = \beta^2 \frac{\partial F_{imp}}{\partial \beta} .
\end{eqnarray}
See Fig. \ref{fig_SpHeatCoeff}, which shows three temperature regimes clearly, corresponding to the decoupled local moment state at high temperatures,
a Kondo fluctuating crossover regime at intermediate temperatures, and the local Fermi-liquid state at low temperatures, respectively. A dip should be
regarded as an artifact of the Schwinger-boson based RG-improved perturbation theory. In order to reveal the origin of this dip behavior, we decompose
the impurity specific heat into two contributions: One comes from the positive-frequency part of the spinon spectral function and the other results from
its negative-frequency sector. The former describes dynamics of the decoupled local moment state and the latter explains the overdamped dynamics of the local
moment due to the Kondo effect, as discussed before. Indeed, dynamics of the decoupled local moment state governs the high temperature regime above $T_{LFL}$,
below which it does not contribute to the specific heat. On the other hand, the overdamped spin dynamics describes the low temperature specific heat below
$T_{LFL}$, above which it does not affect anything. As a result, the dip behavior originates from the fact that dynamics of the decoupled local moment state is
overestimated more than the overdamped spin dynamics. In other words, damping effects should be introduced into the dynamics of the decoupled local moment state,
where its spectrum should be broadened and some spectral weight should be transferred from the negative-energy part. We conclude that the existence of the dip behavior
has the same origin as $S = 1$ for the calculation of the effective spinon chemical potential.

%
%
%
%
%
%
%
%
\begin{figure}[t!]
\includegraphics[width=8cm]{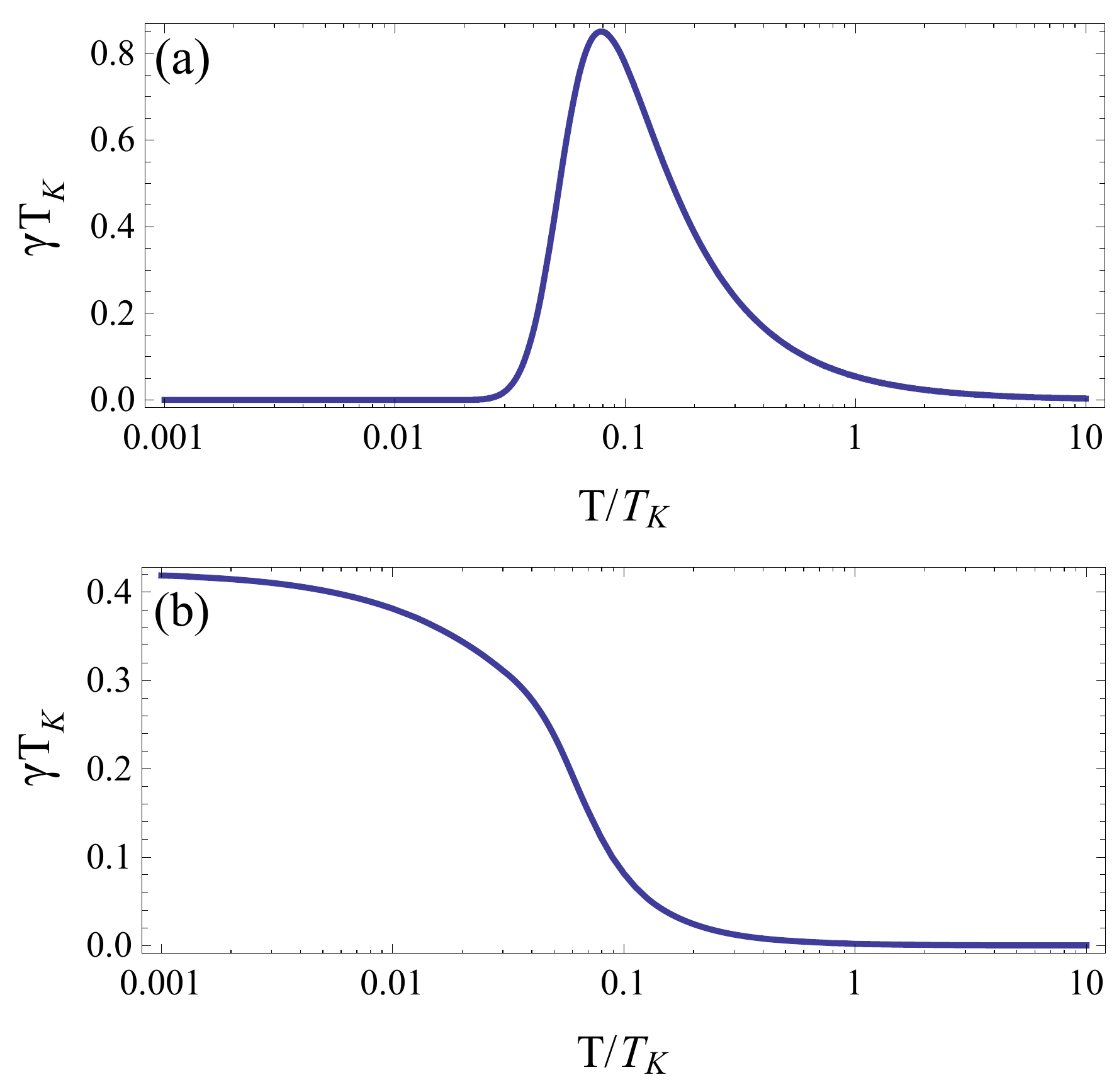}
\caption{ $\gamma$ contributed from (a) the positive- and (b) negative- frequency parts of the spinon spectral function in a log-normal scale. The positive-frequency contribution (a)
may be identified with the specific heat coefficient of the decoupled local moment state while the negative one (b), as that of the overdamped spinon dynamics,
resulting from strong Kondo interactions.
}
\label{fig_SpHeatCoeff_ab}
\end{figure}
%
%

\begin{figure}[t!]
\includegraphics[width=8cm]{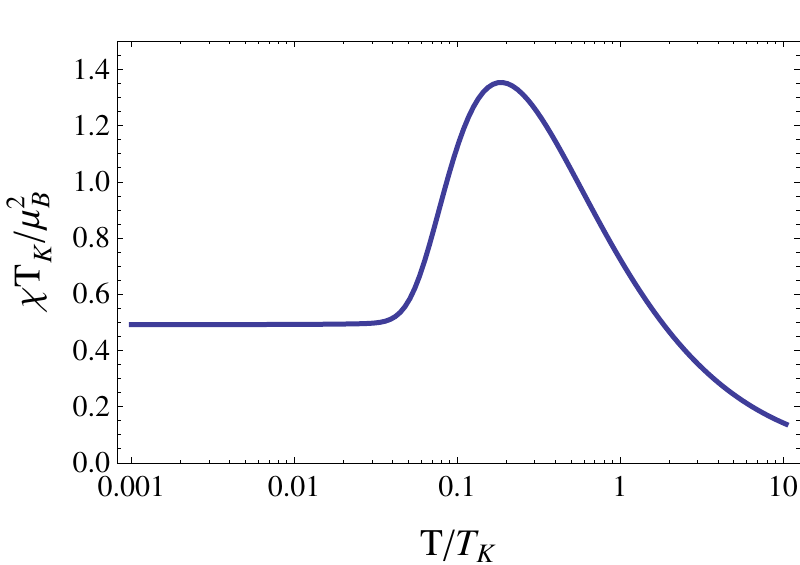}
\caption{Impurity spin susceptibility in a log-normal scale. It shows three temperature regimes clearly, identified with the decoupled local moment state at high temperatures (Curie behavior),
a Kondo fluctuating crossover regime at intermediate temperatures, and the local Fermi-liquid state at low temperatures.
}
\label{fig_SpinSuscpt}
\end{figure}

\begin{figure}[t!]
\includegraphics[width=8cm]{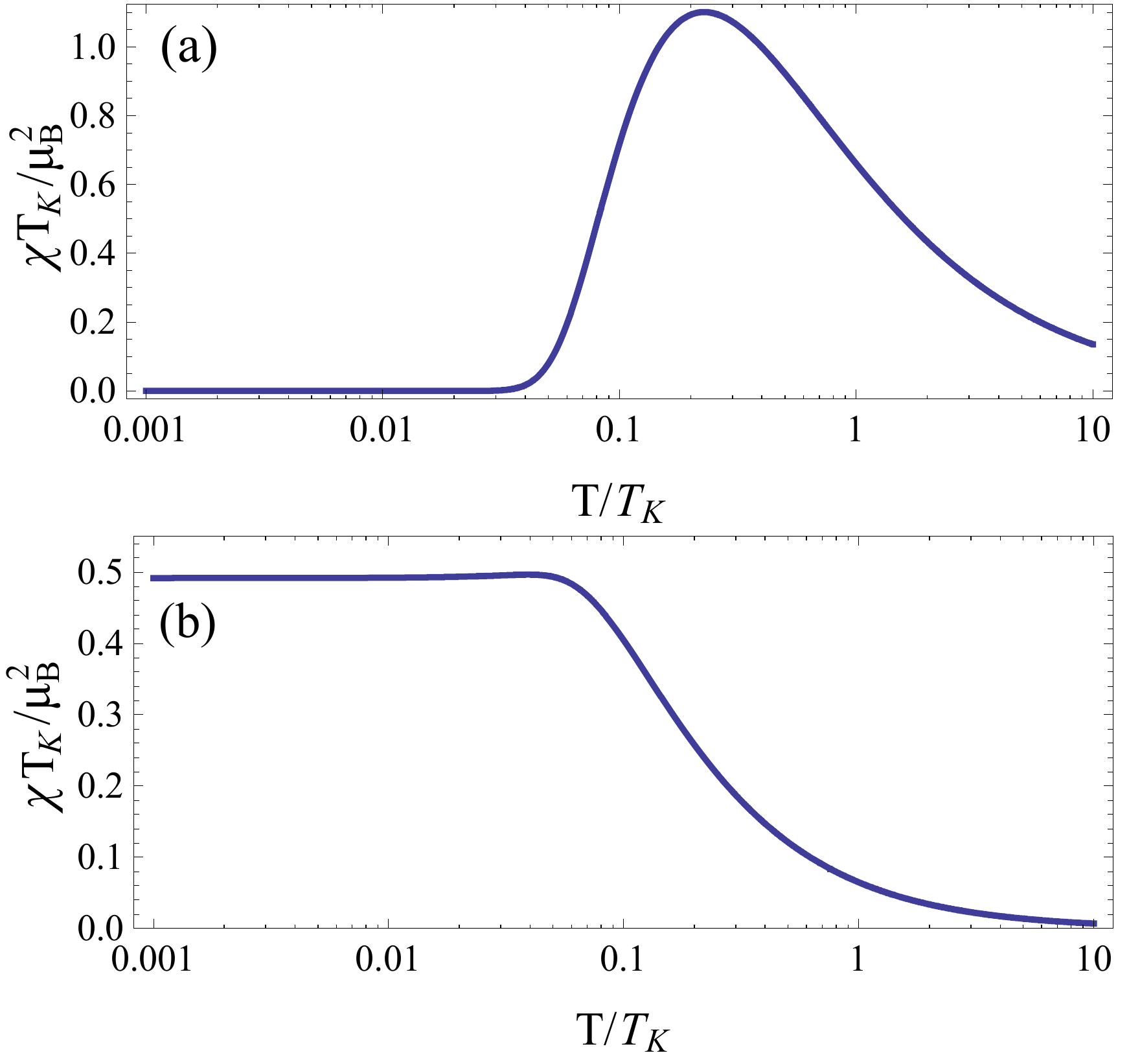}
\caption{
$\chi$ contributed from (a) the positive- and (b) negative- frequency parts of the spinon spectral function in a log-normal scale. The positive-frequency contribution (a)
may be identified with the spin susceptibility of the decoupled local moment state while the negative one (b), as that of the overdamped spinon dynamics,
resulting from strong Kondo interactions.
}
\label{fig_SpinSuscpt_ab}
\end{figure}

In order to find the spin susceptibility of the magnetic impurity, we introduce an effective Zeeman term into the spinon action, given by $S_h = \sum_{i\Omega, \alpha} g\alpha h \  z_{r \alpha}^\dagger (i\Omega) z_{r \alpha}(i\Omega)$, where $g$ is the Lande $g$-factor multiplied by Bohr magneton, $h$ is an applied magnetic field, and $\alpha = \pm 1$ is the spin quantum number. Then, the impurity free-energy $F_{imp}$ including this Zeeman term can be obtained, substituting $\lambda_D$ with $\lambda_D + g \alpha h$. The impurity spin susceptibility is given by
\begin{eqnarray}
\chi = \frac{\partial m}{\partial h}\bigg|_{h=0} = -\frac{2g^2}{\beta} \sum_{i\Omega} \frac{1}{\big[ -i\Omega + \lambda_D + \Sigma_z(i\Omega) \big]^2} ,
\end{eqnarray}
where
\begin{eqnarray}
m = \frac{\partial F_{imp}}{\partial h}
\end{eqnarray}
is magnetization of the impurity. See Fig. \ref{fig_SpinSuscpt}, which also identifies temperature regimes with three as the specific heat coefficient.
In the same way as the impurity specific heat, we analyze the impurity spin susceptibility, decomposing it into the positive-energy and negative-energy contributions
of the spinon spectral weight. As a result, dynamics of the decoupled local moment state contributes to the high temperature regime, responsible for the Curie behavior,
while the overdamped spinon dynamics results in the local-Fermi liquid physics, giving rise to the Pauli-like behavior at low temperatures. See Fig. \ref{fig_SpinSuscpt_ab}.

%
%
%
%
%
%

We would like to emphasize that all these results are consistent with the Schwinger-boson NCA self-consistent theory \cite{Schwinger_Boson_NCA}. There must be reasoning for this consistency between the RG-improved
perturbation theory and the NCA theory. If one compares quantum corrections introduced into the effective theory for both cases, he/she realizes that essentially the same class of Feynman diagrams are summed up.
This is in parallel with the relation between the t-matrix improved perturbation theory and the BCS mean-field theory on superconductivity \cite{BCS_Textbook}, as discussed in the introduction.

\section{Discussions and conclusions}
\label{Discussion}

The Schwinger-boson based RG-improved perturbation theory for the Kondo effect shows two essential features: One is that the spinon self-energy is given by the form of the marginal Fermi-liquid physics,
where its imaginary part is linearly proportional to energy, and the spinon chemical potential is finite at zero temperature, where the statistics of spinons is bosonic.
The marginal Fermi-liquid physics for the spinon self-energy originates from the branch-cut singularity of the renormalized Kondo interaction vertex, which is the solution of the RG equation
for the Kondo interaction coefficient, where its $\beta-$function is negative, typical for the strong coupling problem. The Kondo effect gives rise to such overdamped low-energy dynamics for spinons.
An important point is that both the $\beta-$function or the renormalized Kondo interaction vertex as its solution and the marginal Fermi-liquid physics for the spinon
self-energy do not depend on the statistics of spinons. Both quantities remain the same for either fermion or boson representation of the localized spin. On the other hand, the finiteness of the spinon
chemical potential is allowed only when the statistics of spinons follows that of bosons. The spinon chemical potential vanishes at zero temperature for the slave-boson representation of the localized
spin, resulting in the local non-Fermi liquid fixed point instead of the local Fermi-liquid one. The temperature scale for the local Fermi-liquid state is given by $T_{LFL} \sim 0.1 T_{K}$, where $T_{K}$
is determined by the pole of the renormalized Kondo interaction vertex and $T_{LFL}$ is consistent with the temperature at which the spinon chemical potential starts to become a constant.

The finite spinon chemical potential allows three temperature regimes, corresponding to the decoupled local moment state above $T_{K}$, a strongly fluctuating Kondo regime in $T_{LFL} < T < T_{K}$, and
the local Fermi-liquid state below $T_{LFL}$. Both the specific heat coefficient and spin susceptibility of the magnetic impurity confirm the existence of three temperature ranges. Such thermodynamic quantities
look essentially similar to those of the Schwinger-boson based NCA self-consistent theory \cite{Schwinger_Boson_NCA}. On the other hand, the slave-boson based RG-improved perturbation theory fails to reproduce such three regimes
as the slave-boson based NCA self-consistent theory, the origin of which is that the spinon chemical potential vanishes identically at zero temperature as discussed above. Combined with the slave-boson based CTMA
self-consistent theory, we speculate that vertex corrections of the CTMA theory turn the statistic of spinons from fermions to bosons, giving rise to the local Fermi-liquid state.

Not only the exactly screened Kondo effect but also the over-screened Kondo effect can be described within the RG-improved Schwinger-boson perturbation theory. Keeping the second cubic term in the RG equation (\ref{eq:RGE}), we find a frequency-dependent Kondo coupling function. Resorting to this Kondo vertex, we can describe the non-Fermi liquid fixed point \cite{Kondo_Textbook} within the RG-improved perturbation framework. On the other hand, it is not clear how to reach the singular Fermi-liquid fixed point, which corresponds to the under-screened case \cite{Kondo_Textbook}, where the impurity spin is partially screened through the strong-coupling Kondo vertex and the residual magnetic moment becomes coupled to conduction electrons ferromagnetically. If we focus on the low-temperature regime, we may resort to an RG equation for the ferromagnetic Kondo vertex, allowing us to describe the singular Fermi-liquid state successfully within our perturbation theory. On the other hand, in order to understand the crossover from the local moment state to the singular Fermi-liquid state, we may divide each regime for the RG analysis since the RG equation at high temperatures would differ from that at low temperatures. It is not fully clarified yet the role of the channel number of conduction electrons in the present ``perturbation" theory, more concretely, how the channel number affects the RG analysis \cite{Schwinger_Boson_NCA}.

We would like to claim that the present study implies an interesting and important message for bad metals near Mott quantum criticality, where local magnetic moments may emerge at high temperatures \cite{Bad_Metal_DMFT,Kim_U1SSR}.
In addition, we believe that the Schwinger-boson based RG-improved perturbation theory for the Kondo effect sheds some light on the nature of non-Fermi liquid physics near heavy-fermion quantum criticality,
involved with the mechanism how localized magnetic moments are screened to form a Landau's Fermi-liquid state of heavy electrons \cite{Kim_HF_Schwinger_Boson}.

\section*{Acknowledgments}

This study was supported by the Ministry of Education, Science, and Technology (No. 2012R1A1B3000550 and No. 2011-0030785) of the National Research Foundation of Korea (NRF) and by TJ Park Science Fellowship of the POSCO TJ Park Foundation. We thank Nabyendu Das for initial collaborations. KS appreciates sincere support of Prof. Hyun-Woo Lee.


\section*{References}


\begin{thebibliography}{9}
\bibitem{Kondo original} J. Kondo, . Prog. Theor. Phys. {\bf 32} 37–49 (1964).
\bibitem{Kondo_Textbook} A. C. Hewson, \textit{The Kondo Problem to Heavy Fermions}, (Cambridge University Press, New York, 1993).
\bibitem{NRG_Review} R. Bulla, T. A. Costi, and T. Pruschke, Rev. Mod. Phys. {\bf 80}, 395 (2008).
\bibitem{Bethe_Ansatz} A. O. Gogolin, A. A. Nersesyan, and A. M. Tsvelik, \textit{Bosonization and Strongly Correlated Systems} (Cambridge University Press, New York, 2004).
\bibitem{CTMA_Review} J. Kroha and P. Wolfle, arXiv:cond-mat/0105491: \textit{Theoretical Methods for Strongly Correlated Electrons}, D. Senechal, A.-M. Tremblay, and C. Bourbonnais Eds., CRM Series in Mathematical Physics (Springer, New York, 2003)
\bibitem{Schwinger_Boson_NCA} P. Coleman and C. P\'epin, Phys. Rev. B {\bf 68}, 220405(R) (2003); P. Coleman and I. Paul, Phys. Rev. B {\bf 71}, 035111 (2005); J. Rech, P. Coleman, G. Zarand, and O. Parcollet, Phys. Rev. Lett. {\bf 96}, 016601 (2006); Eran Lebanon and P. Coleman, Phys. Rev. B {\bf 76}, 085117 (2007).
\bibitem{Kim_Tanaka_Review} Ki-Seok Kim and Akihiro Tanaka, arXiv:1412.8254: An invited review article for a special issue on "Skyrmions in Strongly Correlated Systems" in the International Journal of Modern Physics B.
\bibitem{BCS_Textbook} J. R. Schrieffer, \textit{Theory of Superconductivity}, (Westview Press, 1999).
\bibitem{Spin_Textbook} A. Auerbach, \textit{Interacting Electrons and Quantum magnetism} (Springer-Verlag, New York, 1994).
\bibitem{SB_spin liquids} D. P. Arovas and A. Auerbach, Phys. Rev. B {\bf 38}, 316 (1988)
\bibitem{SpN_Read_Sachdev} N. Read and S. Sachdev, Phys. Rev. Lett. {\bf 62}, 1694 (1989); N. Read and S. Sachdev, Phys. Rev. B {\bf 42}, 4568 (1990); S. Sachdev and N. Read, Int. J. Mod. Phys. B {\bf 5}, 219 (1991).
\bibitem{RG_Shankar_RMP} R. Shankar, Rev. Mod. Phys. {\bf 66} 129 (1994).
\bibitem{RG_Textbook} M. E. Peskin and D. V. Schroeder, \textit{An Introduction to Quantum Field Theory} (Addison-Wesley Publishing Company, New York, 1995).
\bibitem{Varma_Review} C. M. Varma, Z. Nussinov, and W. van Saarloos, Phys. Rep. {\bf 361}, 267 (2002).
\bibitem{Bad_Metal_DMFT} X. Deng, J. Mravlje, R. Zitko, M. Ferrero, G. Kotliar, and A. Georges, Phys. Rev. Lett. {\bf 110}, 086401 (2013).
\bibitem{Kim_U1SSR} Ki-Seok Kim, Phys. Rev. B. {\bf 90}, 205129 (2014); Ki-Seok Kim, J. Phys. Soc. Jpn. {\bf 83}, 044709 (2014); Ki-Seok Kim, arXiv:1408.2993; Ki-Seok Kim and Mun Dae Kim, Phys. Rev. B {\bf 77}, 125103 (2008).
\bibitem{Kim_HF_Schwinger_Boson} Ki-Seok Kim and Chenglong Jia, Phys. Rev. Lett. {\bf 104}, 156403 (2010); Ki-Seok Kim and Mun Dae Kim, Phys. Rev. B {\bf 75}, 035117 (2007).

\end{thebibliography}
\end{document}